\documentclass[12pt]{iopart}

\usepackage{graphicx}
\usepackage{iopams}

\def\BE{\begin{equation}}
\def\EE{\end{equation}}
\def\BEA{\begin{eqnarray}}
\def\EEA{\end{eqnarray}}

\newcommand{\fd}[2]{\frac{d #1}{d #2}}

\newcommand{\snr}{\textsf{snr}}

\begin{document}

\title{Carnot in the Information Age:\\ Discrete Symmetric Channels }

\author{Ido Kanter$^1$, Ori Shental$^2$, Hadar Efraim$^1$, Nadav
Yacov$^1$ }

\address{$^1$  Department of Physics, Bar-Ilan University, Ramat-Gan,
52900 Israel}
\address{$^2$ Center for Magnetic Recording Research, University of
California, San Diego 9500 Gilman Drive, La Jolla, CA 92093, USA}

\begin{abstract}
Modeling communication channels as thermal systems results in
Hamiltonians which are an explicit function of the temperature. The
first two authors have recently generalized the second thermodynamic
law to encompass systems with temperature-dependent energy levels,
\mbox{$dQ=TdS+<d\mathcal{E}/dT>dT$}, where \mbox{$<\cdot>$} denotes
averaging over the Boltzmann distribution, recomputing the mutual
information and other main properties of the popular Gaussian
channel. Here the mutual information for the binary symmetric
channel as well as for the discrete symmetric channel consisting of
$4$ input/output (I/O) symbols is explicitly calculated using the
generalized second law of thermodynamics. For equiprobable I/O the
mutual information of the examined channels has a very simple form,
-$\gamma U(\gamma)|_0^\beta$, where $U$ denotes the internal energy
of the channel. We prove that this simple form of the mutual
information governs the class of discrete memoryless symmetric
communication channels with equiprobable I/O symbols.
\end{abstract}

\maketitle

\section{Introduction}
The current scientific conception is that the theory of information
is a creature of mathematics and has its own vitality independent of
the physical laws of nature~\cite{Shannon}. The first two authors
have recently proved~\cite{SMC1,SMC2} that the principal quantity in
the theory of information, the mutual information, can be
reformulated as a consequence of the fundamental laws of nature -
the laws of thermodynamics. This corollary was originally
exemplified for the Gaussian noisy channel.

The generic problem in information processing is the transmission of
information over a noisy
channel~\cite{BookCover,BookBlahut,BookGallager2008}. This central
paradigm of information theory can be mathematically abstracted to
having two random variables \mbox{$X$} and \mbox{$Y$} representing
the desired information and its noisy replica, respectively. Noisy
transmission can occur either via space from one geographical point
to another, as happens in communications, or in time, for example,
when sequentially writing and reading files from a hard disk in the
computer.

Mutual information, \mbox{$I(X;Y)$}, quantifies the amount of
information in common between two random variables and it is used to
upper bound the attainable rate of information transferred across a
channel. To put differently, mutual information measures the amount
of information that can be obtained about one random variable
(channel input \mbox{$X$}) by observing another (output \mbox{$Y$}).
A basic property of the mutual information is that
\mbox{$I(X;Y)=H(X)-H(X|Y)$}, where \mbox{$H(\cdot)$} is the
information (Shannon) entropy~\cite{Shannon}. It measures the amount
of uncertainty in a random variable, indicating how easily data can
be losslessly compressed. Hence knowing \mbox{$Y$}, we can save an
average of \mbox{$I(X;Y)$} bits in encoding \mbox{$X$} compared to
not knowing \mbox{$Y$}.

As recently shown~\cite{SMC1,SMC2}, the modeling of the Gaussian
channel as a thermal system requires the generalization of
thermodynamics to include temperature-dependent Hamiltonians, as
well as the redefinition of the notion of temperature. The
generalized second thermodynamic law was proven to have the
following form
\BE\label{eq_G_second_law}
dS=\frac{dQ}{T}-\frac{1}{T}\bigg<\fd{\mathcal{E}(X)}{T}\bigg>dT
\EE
where $<\cdot>$ denotes averaging over the Boltzmann distribution.

The generalized second law of thermodynamics~(\ref{eq_G_second_law})
has a clear physical interpretation. For simplicity, let us assume
that an examined system is characterized by a comb of discrete
energy levels \mbox{$\mathcal{E}1,\mathcal{E}2,\ldots$}, see Figure
1a . The heat absorbed into the \mbox{$T$}-dependent system has the
following dual effect: A first contribution of the heat,
\mbox{$dU-<d\mathcal{E}(X)/dT>dT$}, increases the temperature of the
system, Figure 1b, while the second contribution,
\mbox{$<d\mathcal{E}(X)/dT>dT$}, goes for shifting the energy comb,
Figure 1c. However, the shift of the energy comb does \emph{not}
affect the entropy, since the occupation of each energy level
remains the same, and the entropy is independent of the energy
values which stand behind the labels
\mbox{$\mathcal{E}1,\mathcal{E}2,\ldots$}. The change in the entropy
can be done only by moving part of the occupation of one tooth of
the energy comb to the neighboring teeth,
Figure 1b. Hence, the {\it effective heat} contributing to the
entropy is \mbox{$dQ-<d\mathcal{E}(X)/dT>dT$}, and this is the
physical explanation to the generalized second
law~(\ref{eq_G_second_law}). A schematic picture of
communication-heat-engine is depicted in Figure~2, where the heat is
devoted to the change of the Hamiltonian (without altering the
thermodynamic entropy) is denoted by the term `working channel'.
Note that for T-independent Hamiltonians the traditional picture of
heat engine is recovered as well as the traditional second
thermodynamic law \cite{Carnot,Reif,Reichl}.

\begin{figure}
\begin{center}
{{\includegraphics[angle=0,scale=0.5]{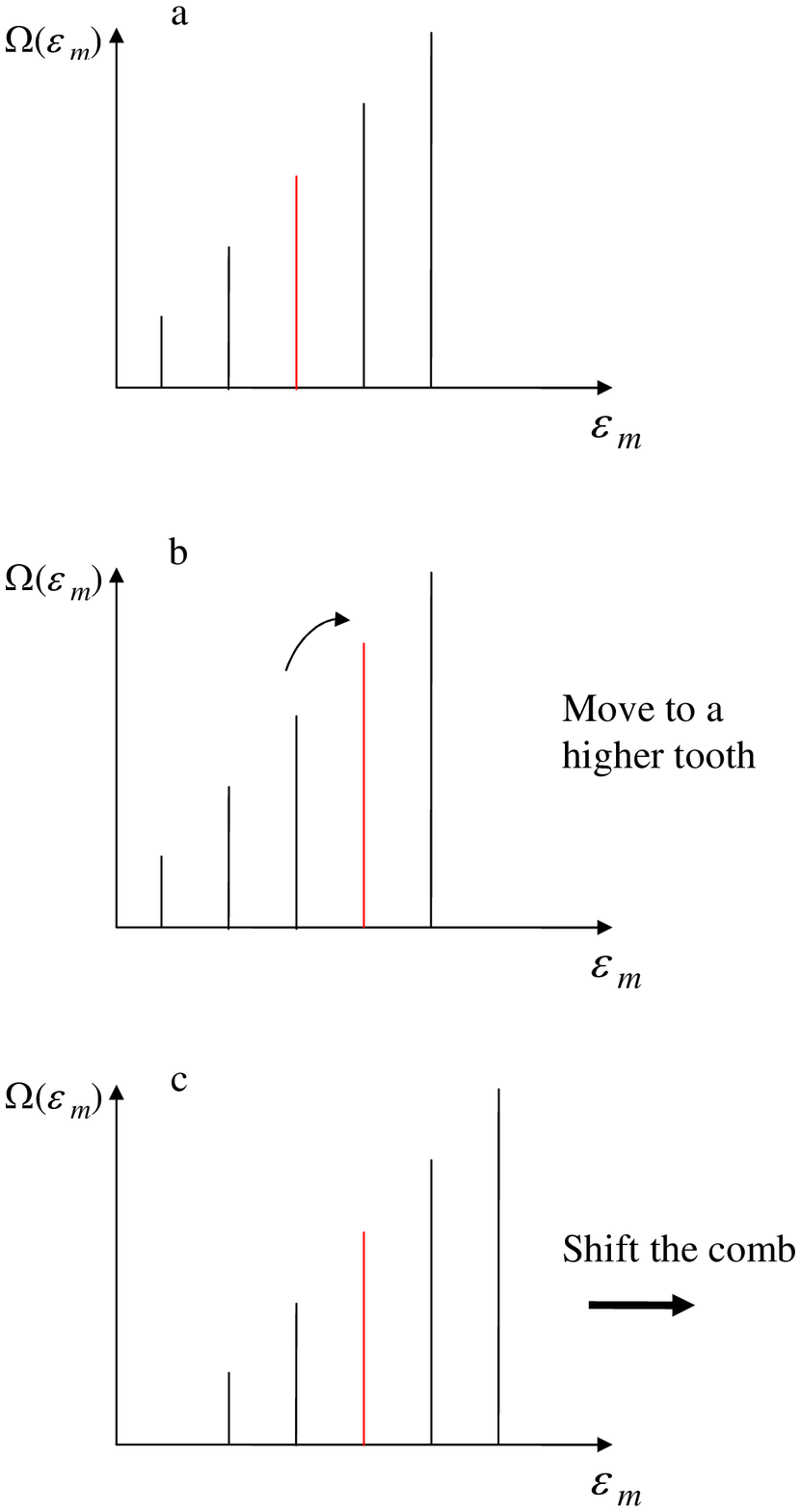}}
\par}
\caption{A system consisting of a discrete energy comb (levels),
$\epsilon_m$, with the corresponding degeneracy $\Omega(\epsilon_m)$
which increases with the energy. For the simplicity of presentation
we assume that the system is occupying only one tooth of the comb,
depicted by a red tooth in Figure 1a. As heat is absorbed into the
system,  the system can either increase the temperature by jumping
to the next tooth of the comb, Figure 1b, or shift the energy comb,
Figure 1c. Note that only the jump to the next tooth, Figure 1b,
changes the entrpoy of the system, where in Figure 1c the entropy
remains the same as in Figure 1a. \label{comb}}
\end{center}
\end{figure}

\begin{figure}
\begin{center}
{{\includegraphics[angle=90,scale=0.3]{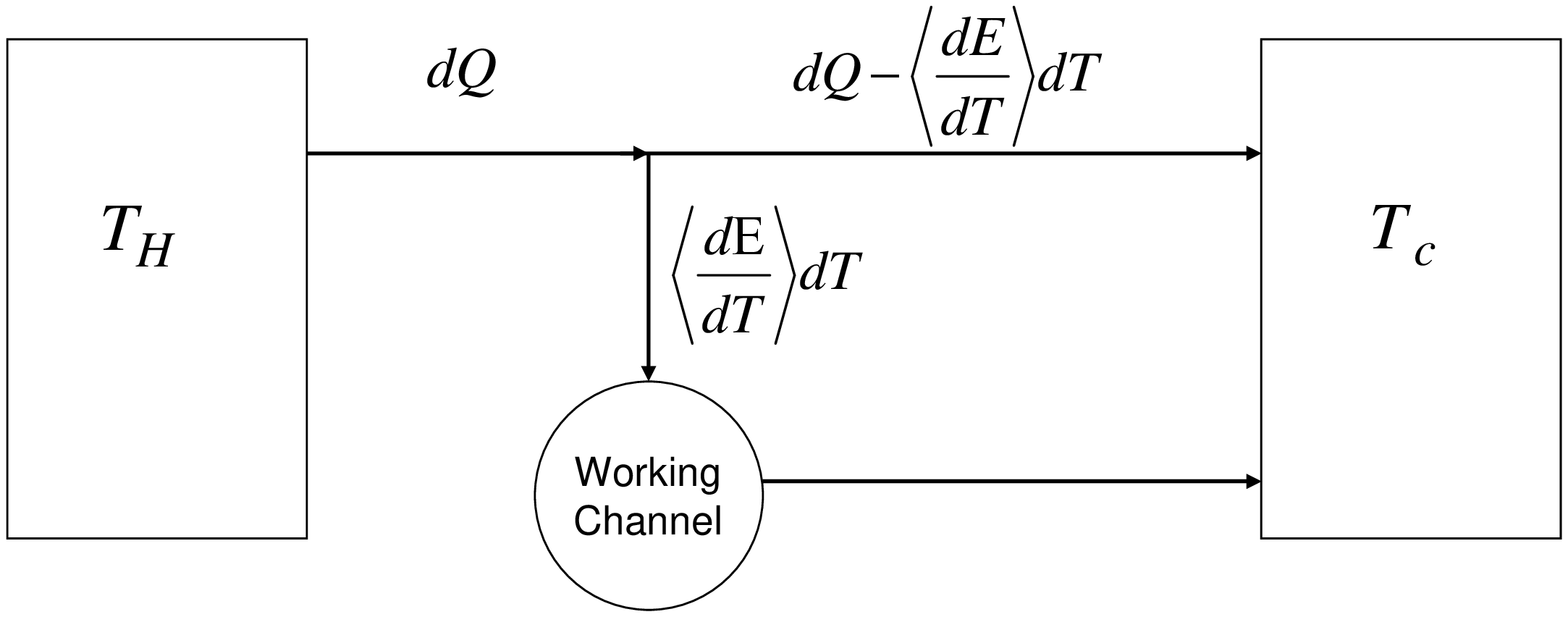}}
\par}
\caption{A schematic communication-heat-engine. H/C denote Hot/Cold
temperatures. \label{enegine}}
\end{center}
\end{figure}

Using the generalized second law of thermodynamics, (1), one can
show~\cite{SMC1,SMC2}that the generalized expression for the mutual
information is given by
\begin{equation} I=-\gamma U(\gamma)|_0^\beta
+\Big<\int_0^{\beta}\Big( U(\gamma,y)+\gamma
<\frac{dE}{d\gamma}>_{x|y} \Big) d\gamma \Big>_y.
\end{equation}
Note that this thermodynamic expression for the mutual information
holds for any channel which can be described by a thermal system
exhibiting quasi-static heat transfer. For the Gaussian channel with
a standard Gaussian input and signal-to-noise ratio, $\snr$, one can
show~\cite{SMC1,SMC2} that the celebrated formula for the Shannon
capacity~\cite{Shannon}, is obtained from (2),
$I(X;Y)=\frac{1}{2}\log{(1+\beta)}$, where $\beta=\snr$.

\section{Binary Symmetric Channel}
We turn now to derive the mutual information for the archetypal
discrete memoryless channel, the binary symmetric channel (BSC),
with input $x$ and output $y$. The input's prior distribution obeys
$P(x=1)=P(x=-1)=1/2$ and the probability for a symbol to flip during
the transmission is denoted by $\delta$, $P(y=\pm|x=\mp)=\delta$,
see Figure 3. Hence, the conditional probability of the output given
the input is
\begin{eqnarray}
P(y|x)=\delta^{\frac{1-xy}{2}} (1-\delta)^{\frac{1+xy}{2}}
=\exp\Big\lbrack
{\frac{xy}{2}\ln(\frac{1-\delta}{\delta})+\frac{1}{2}\ln(\delta(1-\delta))}\Big\rbrack.
\end{eqnarray}
A comparison of the channel's a-posteriori probability distribution,
given by Bayes' law
\begin{equation}
P(X=x|Y=y)\propto P(Y=y|X=x)
\end{equation}

\begin{figure}
\begin{center}
{{\includegraphics[angle=0,scale=0.65]{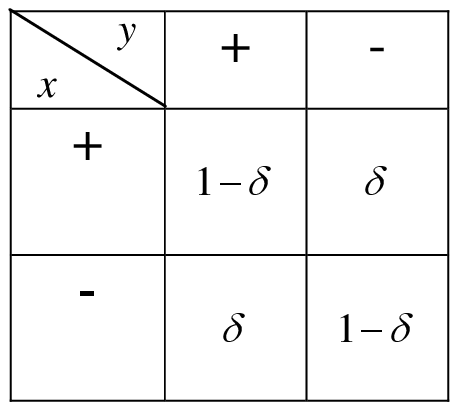}}
\par}
\caption{The input/output transition probabilities for the binary
symmetric channel \label{bsc}}
\end{center}
\end{figure}
with the Boltzmann distribution law yields the following mapping of
the energy and the inverse temperature, $\beta$, of the equivalent
thermal system
\begin{eqnarray}
&& E=-\frac{xy}{2}-\frac{1}{2\beta}\ln(\delta(1-\delta)),
\nonumber \\&&\beta = \ln\frac{1-\delta}{\delta}.
\end{eqnarray}
Note that since the inverse temperature, $\beta(\delta)$, is
positive, then
$\delta=1/(1+\exp(\beta))<1/2$~\cite{Rujan,BookNishimori}. This
definition of $\delta$ coincides with two limiting cases. For
$\delta=1/2$, $\beta=0 ~(T\rightarrow\infty)$, where for $\delta=0$,
$\beta\rightarrow\infty ~(T=0)$. Using $<x|y>=y(1-2\delta)$, one can
find that the internal energy is given by
\begin{equation}
U(\beta)=-\frac{1-2\delta}{2}-\frac{1}{2}+\frac{\ln(1+\exp{(\beta)})}{\beta}.
\end{equation}
\noindent Note that for the BSC specifically the internal energy is
independent of $y$ due to the binary nature of the I/O symbols.
Similarly to the internal energy on can find
\begin{equation}
<\beta \frac{dE}{d\beta}>_{x|y}
=-\frac{\ln(1+\exp{(\beta)})}{\beta}+\frac{\exp{(\beta)}}{1+\exp{(\beta)}}.
\end{equation}
It is easy now to verify that the second term of the generalized
mutual information, (2), vanishes
\begin{equation}
U(\beta)+<\beta \frac{dE}{d\beta}>_{x|y}=
-\frac{\ln(1+\exp{(\beta)})}{\beta}
+\frac{\exp{(\beta)}}{1+\exp{(\beta)}}=0,
\end{equation}
\noindent and the mutual information for the equiprobable
input-output (I/O) BSC (which is also the Shannon capacity of the
BSC in general) has a very simple form given explicitly by
\begin{eqnarray}
&&I=-\gamma U(\gamma)|_0^\beta=\beta(1-\delta)-\ln(1+\exp{(\beta)})+1 = \nonumber \\
&&(1-\delta)\ln(1-\delta)+\delta \ln(\delta)+1 = 1-H(\delta)~.
\end{eqnarray}
\noindent  Hence, the mutual information for the BSC is recovered
using the generalized second thermodynamic law. Note that in
contrary to the Gaussian channel, in the BSC case the second term of
the mutual information, (2), vanishes, and the mutual information is
proportional to the internal energy
\begin{equation}
I=-\gamma U(\gamma)|_0^\beta.
\end{equation}

\section{4-I/O Symbols}
Is this simple thermodynamic form, (10), a coincidence of the BSC
only or it occurs for the general class of discrete memoryless
channels? To find the answer to this interesting question, we first
turn to discuss in detail the case of $4$-I/O symbols. The input is
represented by two binary units $x_1$ and $x_2$, and similarly the
output units are $y_1$ and $y_2$. We assume that the 4-input symbols
are equiprobable. The conditional probability of
$P(y_1,y_2|x_1,x_2)$ obeys the following symmetry. The probability
that both output units are equal to the input units, respectively,
is $1-\delta$, the probability that only one unit is equal is
$\epsilon \delta$  ($\epsilon\le 1$), and the probability that the
two output units differ from the input is $\delta-2\epsilon \delta$,
Figure 4. Hence,
\begin{equation}
\!\!\!\!\!\!\!\!\!\!\!\!\!\!\!\!\!\!\!\!\!\!\!\!\!\!\!\!
P(y|x)=(1-\delta)^{\frac{1+x_1y_1}{2}\frac{1+x_2y_2}{2}} (\epsilon
\delta)^{\frac{1+x_1y_1}{2}\frac{1-x_2y_2}{2}} (\epsilon
\delta)^{\frac{1-x_1y_1}{2}\frac{1+x_2y_2}{2}}
(\delta-2\epsilon\delta)^{\frac{1-x_1y_1}{2}\frac{1-x_2y_2}{2}},
\end{equation}
\noindent and it is easy now to verify that
\begin{eqnarray}
\!\!\!\!\!\!\!\!\!\!\!\!\!\!\!\!\!\!\!\!\!\!\!\!\!\!\!\!\!\!\!\!\!\!\!\!
&&P(X=x|Y=y) \propto \nonumber
\\\!\!\!\!\!\!\!\!\!\!\!\!\!
\!\!\!\!\!\!\!\!\!\!\!\!\!\!\!\!\!\!\!\!\!
&&\exp{\Big\{\frac{x_1y_1+x_2y_2}{4}\ln(\frac{1-\delta}{\delta(1-2\epsilon)})+\frac{1}{4}
\ln((1-\delta)\delta^3\epsilon^2(1-\epsilon))
+\frac{x_1x_2y_1y_2}{4}\ln\lbrack
\frac{1-\delta}{\delta}\frac{(1-2\epsilon)}{\epsilon^{2}}\rbrack\Big\}}.\nonumber
\end{eqnarray}
\begin{figure}
\begin{center}
{{\includegraphics[angle=0,scale=0.75]{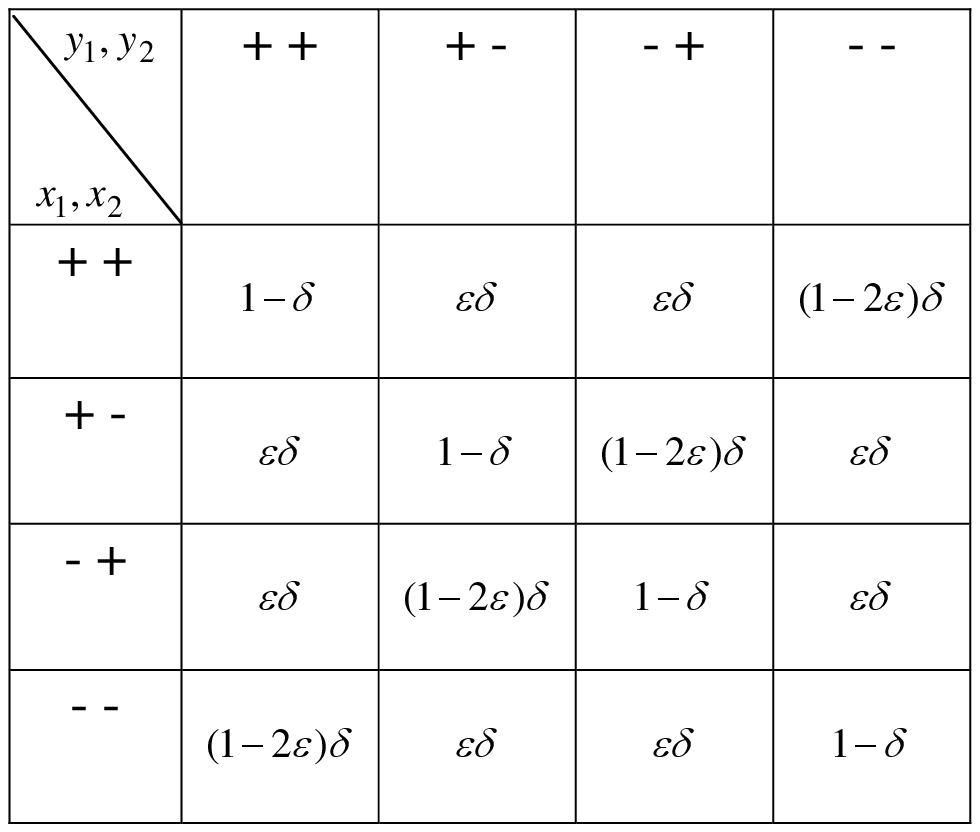}}
\par}
\caption{The input/output transition probabilities for a channel of
4-I/O symbols. \label{matrix_4}}
\end{center}
\end{figure}
\noindent Similarly to (5) one can find in this case
\begin{eqnarray}
\!\!\!\!\!\!\!\!\!
E=-\frac{1}{4}(x_1y_1+x_2y_2+x_1x_2y_1y_2)-\frac{1}{4}\ln((1-\delta)\delta^3)
+\frac{1}{4\beta}f(x_1y_1,x_2y_2,\epsilon),
\end{eqnarray}
\noindent where
\begin{equation}
f\triangleq(x_1y_1+x_2y_2)\ln(1-2\epsilon)-x_1x_2y_1y_2\ln(\frac{1-2\epsilon}{\epsilon^2})
-\ln(\epsilon^2(1-\epsilon))~,
\end{equation}
\begin{equation}
\beta=\ln(\frac{1-\delta}{\delta})~,
\end{equation}
and the two limiting cases are $\delta=3/4$ and $\epsilon=1/3~
(\beta=0)$ and $\delta=0 ~(\beta\rightarrow\infty)$. Using the
following conditional expectations
\begin{eqnarray}
&&<x_m|y_m>=y_m(1-2\delta+2\epsilon\delta)=y_m(1-2\delta(1-\epsilon)), \nonumber\\
&&<x_1x_2|y_1y_2>=y_1y_2(1-4\epsilon\delta),
\end{eqnarray}
one can find that the internal energy
\begin{equation}
U(\beta)=-\frac{3}{4}+\delta-\frac{\ln((1-\delta)\beta)}{4\beta}+\frac{\delta}{\beta}\lbrack
-2\epsilon \ln(\epsilon)+\ln(1-2\epsilon)\rbrack~,
\end{equation}
and again verify that
\begin{equation}
U(\beta)+\beta<\frac{dE}{d\beta}>_{x|y}=0.
\end{equation}
\noindent Hence the mutual information has again the simple form,
(10), and is given explicitly by
\begin{equation}
I=2+(1-\delta)\ln(1-\delta)+2\epsilon\delta \ln(\epsilon\delta)
+\delta(1-2\epsilon)\ln((1-2\epsilon)\delta),
\end{equation}
\noindent which can be verified (by direct computation of the mutual
information out of its definition) to have the correct form. The
mutual information as a function of $(\delta,\epsilon)$ is depicted
in Figure 5, where the mutual information $I=0$ for $\delta=3/4$ and
$\epsilon=1/3$.
\begin{figure}
\begin{center}
{{\includegraphics[angle=0,scale=0.4]{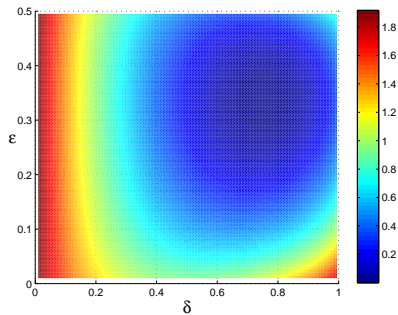}}
\par}
\caption{The mutual information as a function of $\delta$ and
$\epsilon$ for the discussed case of 4-I/O symbols. \label{}}
\end{center}
\end{figure}

\section{Equiprobable Discrete Memoryless Symmetric Channels}
For the general case of discrete
memoryless symmetric channels with $2^n$ equiprobable input symbols,
the conditional probabilities has the following form
\begin{equation}
P(y_1,~...,y_{n}|x_1,~...,x_{n})=\delta
\epsilon(x_1y_1,~...,~x_{n}y_{n}),
\end{equation}
\noindent and one can verify that similarly to (5) and (12) the
energy is given now by
\begin{eqnarray}
E=&& -\frac{1}{2^n}\lbrack -1+\Pi_{k=1}^{n}(1+x_ky_k)\rbrack
-\frac{1}{2^n}\ln((1-\delta)\delta^{2^n-1})\nonumber \\
&&+\frac{1}{\beta}f_n(\{x_iy_i\},~\{\epsilon(x_1y_1,~...,~x_{n}y_{n})\}),
\end{eqnarray}
\noindent where $f_n$ is a function of the number of symbols only.
Now it is clear that
\begin{eqnarray}
\!\!\!\!\!\!\! \!\!\!\!\!\!\!\!\!\!\!\!\!\!\!\!\!\!\!\!\! \!\!\!\!\!
2^n(U(\beta)+<\beta \frac{dE}{d\beta}>_{x|y})=
-<-1+\Pi_{k=1}^{n}(1+x_ky_k)>-\frac{d
\ln((1-\delta)\delta^{2^n-1})}{d\delta}\frac{d\delta}{d\beta},\nonumber
\end{eqnarray}
\noindent where as in (5) and (12) $\beta=\ln((1-\delta)/\delta)$
and $d\delta/d\beta =-\delta(1-\delta)$. Hence
\begin{eqnarray}
\!\!\!\!\!\!\!\!\!\!\!\!\!\!\!\!\!\!2^n(U(\beta)+<\beta
\frac{dE}{d\beta}>_{x|y})= &&
-<-1+\Pi_{k=1}^{n}(1+x_ky_k)>+2^n-1-2^n\delta= \nonumber\\&&
-<\Pi_{k=1}^{n}(1+x_ky_k)>+2^n(1-\delta)=0.
\end{eqnarray}
The identity to $0$ is a result of $<\Pi_{k=1}^{n}(1+x_ky_k)>$ which
is equal to the trace of the conditional probability, (19). Hence we
proved that for a general memoryless symmetric channel consisting of
$2^n$ equiprobable I/O symbols
\begin{eqnarray}
&&I=-\gamma U(\gamma)|_0^\beta, \nonumber \\
&&U(\beta)+<\beta \frac{dE}{d\beta}>_{x|y}=0.
\end{eqnarray}

Note that the identity $I= -\gamma U(\gamma)|_0^\beta$ is in
contrast to the case of a Gaussian channel with Bernoulli-1/2 or
Gaussian inputs~\cite{SMC1,SMC2}.



\end{document}